# --- Cover page ---

# Disorder and cavity evolution in single-crystalline Ge during implantation of Sb ions monitored in-situ by spectroscopic ellipsometry

Tivadar Lohner[1], Attila Németh[2], Zsolt Zolnai[1], Benjamin Kalas[1], Alekszej Romanenko[1], Nguyen Quoc Khánh[1], Edit Szilágyi[2], Endre Kótai[2], Emil Agócs[1], Zsolt Tóth[3], Judit Budai[4,5], Péter Petrik[1,*], Miklós Fried[1,6], István Bársony[1], and †József Gyulai[1]

[1]Institute of Technical Physics and Materials Science, Centre for Energy Research, Konkoly-Thege M St. 29-33, 1121 Budapest, Hungary

[2]Institute for Particle and Nuclear Physics, Wigner Research Centre for Physics, Konkoly-Thege M St. 29-33, 1121 Budapest, Hungary

[3]Department of Medical Physics and Informatics, University of Szeged, Korányi fasor 9., H-6720 Szeged, Hungary

[4]ELI-ALPS, ELI-HU Non-Profit Ltd., Wolfgang Sandner St. 3., 6728 Szeged, Hungary

[5]Department of Optics and Quantum Electronics, University of Szeged, Dóm tér 9, 6720 Szeged, Hungary

[6]Institute of Microelectronics and Technology, Óbuda University, P.O. Box 112, 1431 Budapest, Hungary

*Corresponding author

**Acknowledgements**

The authors are grateful for financial support from the OTKA K131515 and K129009 projects. This project (20FUN02"POLight") has received funding from the EMPIR programme co-financed by the Participating States and from the European Union's Horizon 2020 research and innovation programme.

# Disorder and cavity evolution in single-crystalline Ge during implantation of Sb ions monitored in-situ by spectroscopic ellipsometry

Tivadar Lohner[1], Attila Németh[2], Zsolt Zolnai[1], Benjamin Kalas[1], Alekszej Romanenko[1], Nguyen Quoc Khánh[1], Edit Szilágyi[2], Endre Kótai[2], Emil Agócs[1], Zsolt Tóth[3], Judit Budai[4,5], Péter Petrik[1,*], Miklós Fried[1,6], István Bársony[1], and †József Gyulai[1]

[1]Institute of Technical Physics and Materials Science, Centre for Energy Research, Konkoly-Thege M St. 29-33, 1121 Budapest, Hungary

[2]Institute for Particle and Nuclear Physics, Wigner Research Centre for Physics, Konkoly-Thege M St. 29-33, 1121 Budapest, Hungary

[3]Department of Medical Physics and Informatics, University of Szeged, Korányi fasor 9., H-6720 Szeged, Hungary

[4]ELI-ALPS, ELI-HU Non-Profit Ltd., Wolfgang Sandner St. 3., 6728 Szeged, Hungary

[5]Department of Optics and Quantum Electronics, University of Szeged, Dóm tér 9, 6720 Szeged, Hungary

[6]Institute of Microelectronics and Technology, Óbuda University, P.O. Box 112, 1431 Budapest, Hungary

*Corresponding author

## ABSTRACT

Ion implantation has been a key technology for the controlled surface modification of materials in microelectronics and generally, for tribology, biocompatibility, corrosion resistance and many more. To form shallow junctions in Ge is a challenging task. In this work the formation and accumulation of shallow damage profiles was studied by in-situ spectroscopic ellipsometry (SE) for the accurate tracking and evaluation of void and damage fractions in crystalline Ge during implantation of 200-keV $Sb^+$ ions with a total fluence up to $10^{16}$ $cm^{-2}$ and an ion flux of $2.1 \times 10^{12}$ $cm^{-2}s^{-1}$. The consecutive stages of damage accumulation were identified using optical multi-layer models with quantitative parameters of the thickness of modified layers as well as the volume fractions of amorphized material and voids. The effective size of damaged zones formed from ion tracks initiated by individual bombarding ions can be estimated by numerical simulation compared with the dynamics of damage profiles measured by ion beam analysis and ellipsometry. According to our observations, the formation of initial partial disorder was followed by complete amorphization and void formation occurring at the fluence of about $1 \times 10^{15}$ $cm^{-2}$, leading to a high volume fraction of voids and a modified layer thickness of ≈200 nm by the end of the irradiation process. This agrees with the results of numerical simulations and complementary scanning electron microscopy (SEM) measurements. In addition, we found a quasi-periodic time dependent behavior of amorphization and void formation represented by alternating accelerations and decelerations of different reorganization processes, respectively. For the understanding and prevention of adverse void formation and for controlled evolution of subsurface nanocavities or cellular surface texture the in-situ monitoring of the dynamics of structural damage accumulation by the developed SE method is essential.

1. Introduction

The interest in Ge has been increasing in recent years due to the broad range of potential applications from photonics [1] to microelectronics. It has been utilized in high-speed photodetectors that are compatible with Si microtechnology, thus providing a cost-effective alternative to III-V semiconductors [2]. Ge and its alloys (primarily with Si) are also used as Bragg reflectors [3], photodiodes [2], light-emitting layers (photo- and electroluminescence, injection lasers) [4,5], materials of controlled optical properties (especially in the infrared wavelength range) [6], as well as in band gap [7] and refractive index [8] engineering. Although the size-dependent optical properties of Ge were measured earlier [9] than those of Si [10], the attention of researchers has turned only recently to Ge nanocrystals [11,12].

In the fabrication of device structures ion implantation is used for the controlled doping or modification of crystallinity [13], due to the fact that both energy and fluence of the ion beam can be precisely controlled, and the material properties obtained by ion implantation are highly reproducible. Furthermore, the temporal and spatial variation of disorder can be influenced by the choice of the implanted element [14] and the ion flux, i.e., the ion current [15]. The profile of the implanted elements and disorder can be controlled by the energy [16,17] and direction of the implanted ions [18].

Depending on which material property was in the focus of the study, various methods of characterization of the implanted Ge structure had been applied. Structural measurements are usually conducted by high-resolution scanning [19] or transmission [8] electron microscopy, HR-SEM and HR-TEM, respectively. Both surface topography [19] and in-depth microstructure [8] have been imaged, revealing a sponge-like structure with voids in many cases, e.g., for 50-300-keV self-ion bombardment [19]. Structural properties such as the size of crystal grains or the degree of amorphization can also be determined by optical methods [8,9,11,20]. Rutherford backscattering spectrometry (RBS) allows revealing the degree of crystallinity, crystal structure (when combined with channeling) and the depth distribution of elements. Its combination with spectroscopic ellipsometry (SE) is a powerful way of complementary depth profiling, primarily in crystalline semiconductors [21–24]. Besides determination of the optical properties [20,25–27], frequently in SiGe alloys [8,22,28], used in optics or optoelectronics, optical methods are also capable of indirectly determining material properties, such as the crystallinity [16], disorder [29] or crystal grain size.

SE [30] is one of the most sensitive methods to detect the change of crystallinity in semiconductors. The measuring technique determines the complex dielectric function of $\varepsilon = \varepsilon_1 + i\varepsilon_2$, $\varepsilon_1$ and $\varepsilon_2$ being the real and imaginary parts, respectively, with a precision of ≈$10^{-4}$. Since $\varepsilon_2$ is proportional to the joint density of electron states, the method is very sensitive to the change of long-range order in the lattice, i.e., to the crystallinity of the material [31]. Due to the large amount of spectroscopic data, complex models can be built with parameters that describe the formation of structures in depth [24,32]. Owing to the non-destructive nature of SE, in-situ applications were demonstrated by many authors [33]. Albeit many papers discussed ex-situ ellipsometric characterization of Ge structures [25], in-situ measurement of ion implantation has only been performed upon cleaning and layer removal by He [34] and Ar [35] ions. Hu et al. characterized the etching and damage evolution in the top 10 nm of native and oxidized Si wafers during implantation with low energy (0.3-1.2 keV) Ar [35] ions by in-situ SE. In our case, as the initial phase of amorphization occurs within a couple of seconds, it can only be followed and analyzed if complete ellipsometric spectra can quickly be recorded. A further advantage of the in-situ approach is that in-vacuum conditions decrease the amount of water molecules and hydrocarbon contamination [36] at the surface, although ex-situ ellipsometry can take into account the modified surface [14,24]. Numerous complementary investigations have revealed the reliability of SE verified by SEM and RBS [24,37].

A number of studies are dealing with the optical and structural characterization of Ge prepared by a wide range of methods [38,25,26,20], but only a few papers discuss the optical and structural characterization of disordered Ge layers obtained by ion bombardment [16]. SE measurements performed on ultra-thin (less than 10 nm) ion implantation-amorphized Ge (i-a-Ge) layers made it possible to determine the dielectric function of i-a-Ge with more than 10% uncertainty. Broad damage distributions were realized by multiple-energy implantations, to made benchmark optical references [20].

Implantation of light elements into c-Ge was used by several authors to create a dense and uniform i-a-Ge structure in a controlled way [16,39]. Previously we used double-energy Al-implantation to create an amorphous layer, which was a suitable reference for the effective medium approximation (EMA) model (see also Fig. S5 in the Supplementary Material). When implanting heavy elements in Ge (Ge ions or heavier species), void formation occurs, and above a certain fluence a peculiar cellular structure develops [8,16,17,19,39–44]. Similar effects were observed in Si [45], CdTe [46] and SiC [47]. Damage and void formation in Ge were investigated by several groups, but no in-situ measurements have been performed. All studies so far have been carried out after the implantation process, and only the final structure have been characterized after reorganization and relaxation processes. To our knowledge there is only a single study [35] in which the layer structure has optically been measured during the implantation process. Furthermore, there is no available results of in-situ SE measurement performed in vacuum for Ge ion implantation.

The aim of this work is to describe the dynamics of the complex process of amorphization and void formation in the technologically important case of shallow ion implantation into Ge. For this, Ge was implanted by heavy Sb ions, and the time evolution of implantation-induced disorder and morphology changes were followed by in-situ SE measurements. Firstly, an ellipsometer onto an ion implantation chamber was mounted, and complete spectra were recorded in the visible-near ultraviolet wavelength range in each 3 s during the implantation of 200-keV Sb ions. In our previous work [20], the complex dielectric function of i-a-Ge was determined by SE thus providing useful input for the evaluation of the present in-situ measurements. By fitting the parameters of a complex optical model we follow the time evolution of the implanted structure changes, such as the thicknesses of heavily and slightly modified sublayers, and the volumetric ratios of c-Ge, i-a-Ge and voids.

## 2. Experimental methods

### 2.1 Sample preparation

Ge wafer from Umicore (orientation (100), resistivity of approx. 0.4 Ω·cm) was cleaned in diluted HF and rinsed in deionized (DI) water. The Ge wafer has a resistivity of approx. 0.4 Ω·cm corresponding to about $5\times10^{15}$ cm$^{-3}$ doping concentration for the n-type material used in this study. After cutting into small rectangular pieces, the samples were rinsed again in DI and dried in $N_2$ gas. The 200-keV $Sb^+$ ions were implanted into a c-Ge piece at a fluence of $1\times10^{16}$ cm$^{-2}$ and ion flux of $2.1\times10^{12}$ cm$^{-2}$s$^{-1}$, corresponding to a current density of 0.34 µA.

### 2.2 Integration of SE into a vacuum chamber

For real-time in-situ SE measurements an M-88 spectroscopic ellipsometer (J.A. Woollam Co. Inc.) with a rotating analyzer was mounted on a high vacuum chamber of the Heavy Ion Cascade Implanter of the Hungarian Ion-beam Physics Platform at the Institute for Particle and Nuclear Physics, Wigner Research Centre for Physics, Budapest. The vacuum chamber shown in Fig. 1 is equipped with high-quality entrance and exit windows to minimize the deviation in the polarization state of the measuring light caused by birefringence. The angle of incidence is fixed at 75°. The spectral range from 367 to 746 nm was measured in 72 spectral points in every 3 seconds, simultaneously. The spectra in the UV and IR regions were limited by the windows and the instrument, respectively. The alignment of the sample was realized by a 4-axis precision goniometer to maximize the reflected intensity, followed by an offset calibration. The sample holder includes a Faraday cup that reduces the typical value of uncertainty of the ion current measurement from 5-10% to 1% [48].

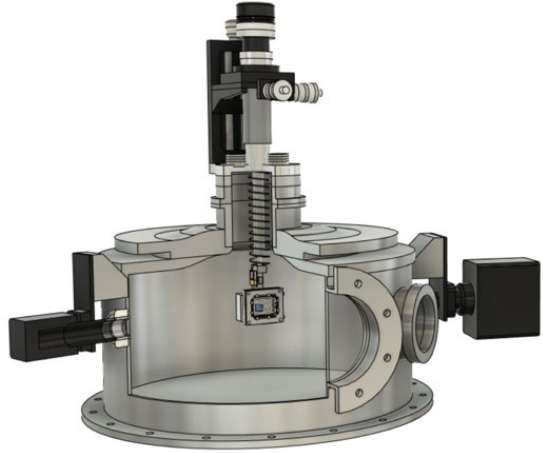

**Figure 1.** Integration of a Woollam M88 spectroscopic ellipsometer in a vacuum chamber. The high-precision goniometer is mounted on the top of the chamber. Black boxes show the arms of the ellipsometer. The sample is mounted on a 4-axis precision goniometer equipped with a Faraday cup for ion current measurement.

**2.3 Measurement by ex-situ SE**

For ex-situ ellipsometric characterization, an M-2000DI (Institute of Technical Physics and Materials Science, Centre for Energy Research, Budapest, Hungary) and an M-2000F (Department of Optics and Quantum Electronics, University of Szeged, Hungary) rotating compensator spectroscopic ellipsometer was used. The effect of a possible surface contamination layer (<1 nm, [36] which should be formed due to the fact that the measurements were not performed in vacuum) is taken into account as part of the surface roughness layer. The M-88 spectroscopic ellipsometer attached to the implantation chamber was operated in the wavelength range from 367 to 746 nm.

**2.4 Evaluation of SE spectra**

SE measures spectra of $\Psi$ and $\Delta$, that describes the complex reflectance ratio of $\rho = r_p/r_s = \tan(\Psi) \cdot \exp(i\Delta)$, where $r_p$ and $r_s$ denote the complex reflection ratio of light polarized parallel and perpendicular to the plane of incidence, respectively [30,49]. Pseudo refractive index (n) and extinction coefficient (k) can be analytically calculated from $\Psi$ and $\Delta$, supposing reflection from a planar surface of a bulk homogeneous medium. The response of the real sample upon reflection of light is calculated using the transfer matrix method based on stratified media with modeled dispersion and thicknesses. The parameters are searched in the global parameter space, then fitted using the Levenberg-Marquardt gradient method to find the smallest discrepancy between the simulated and measured $\Psi$ and $\Delta$ spectra. The quality of the fit is measured by the root mean square error (MSE) between those spectra.

For the construction of optical models for evaluation of the SE spectra we need to determine the reference dielectric function of ion implantation amorphized Ge. The reference amorphous layer has been created by two step amorphization (120 keV Al$^+$ 1x10$^{16}$ atom/cm$^2$ and 300 keV Al$^+$ 1x10$^{16}$ atom/cm$^2$). The spectra obtained from multiple-angles-of-incidence spectroellipsometric measurements were evaluated using a two-layer optical model. The WVASE32, CompleteEASE 6.41 as well as our self-developed software were used for the evaluation of the measured spectra. Reference dielectric data for ion-implantation amorphized Ge has been created from the results of SE evaluation published in [20]. The complex dielectric function of the amorphized Ge layer was described in this work by the Tauc-Lorentz dispersion relation using the above reference for the initial parameter set, fitting only the energy position of the Lorentz oscillator for i-a-Ge. The (MSE) of the fit was between 10-20 over the whole process. The typical uncertainty of the determination of layer thicknesses and volume fractions was a few nanometers and a few percent, respectively.

**2.5 Verification and complementary methods**

The construction and verification of an optical model is the most important step in SE data analysis. In the supplementary material we show two examples which are very similar to our case. Vedam et al. [50] used TEM to verify the thicknesses of a-Si layers created by Si ions (self-implantation) measured by SE (Fig. S1), whereas in our previous work it has been shown that cavity profiles in Si created by high-dose He implantation and annealing measured by SE agree very well with the profiles determined by TEM (Fig. S2 and Ref. [37]). In the present work our optical model was verified with our previous investigation on double-step, double-energy Al-implantation to create an amorphous reference dielectric function for the EMA method. EMA models were successfully applied for the characterization of damage profiles in Ge [20], which show a significant difference („fingerprint") from the dielectric function of c-Ge and voids.

In this study, the optical measurements were completed by high resolution SEM analysis and SRIM (Stopping and Range of Ions in Matter [51]) simulations. We also compared our optical model with the XTEM results of Darby et al. [42] which are shown also in the Supplementary Material (Figs. S3 and S4).

**3. Results**

**3.1 In-situ SE measurement during ion implantation**

The evolution of the derived pseudo refractive index (*n*) and extinction coefficient (*k*) spectra is plotted in Fig. 2. The ion beam was turned on at a time of ≈60 s (a sharp feature in the recorded spectra close to the zero fluence position), and it was turned off at a fluence of 1x10$^{16}$ cm$^{-2}$ (at a time of ≈4800 s). The continuous evolution of all the spectra is evident from Fig. 2, underlining the need for in-situ SE measurement. Note the rapid changes at the start and at the fluence of ≈2x10$^{15}$ cm$^{-2}$. The optical penetration depth (OPD) shows these characteristic features for the red part of the spectrum (above the wavelength of ca. 600 nm), as the penetration depth abruptly decreases at the beginning of the irradiation due to amorphization and increases again in the above-mentioned range between 1.9 and 2.5x10$^{15}$ cm$^{-2}$ due to void formation, as it will be shown in the detailed optical model analysis below.

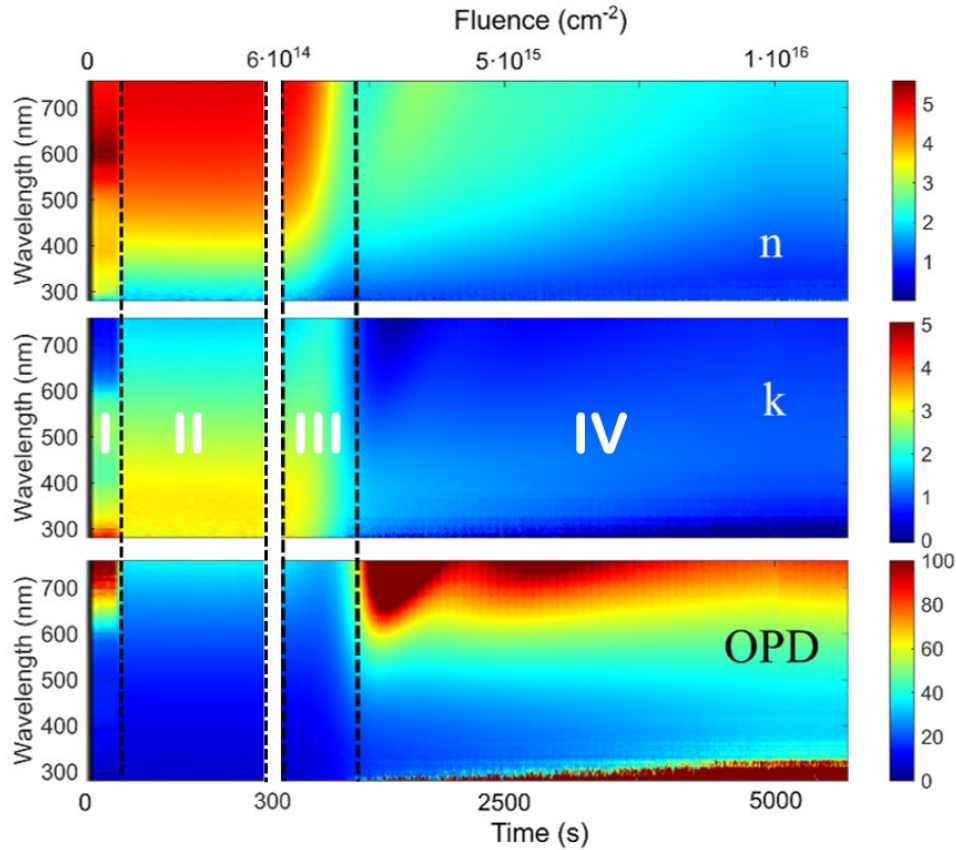

**Figure 2.** Pseudo n, k and pseudo optical penetration depth (OPD = $\lambda/(4\pi k)$, where $\lambda$ denotes the wavelength of light in vacuum) values vs. time derived from in-situ SE measurements performed during the implantation of $Sb^+$ ions into c-Ge. The OPD values have singularity when $k \to 0$, therefore, the OPD is plotted only up to the depth of 100 nm. The vertical dashed lines show the boundaries of the regions discussed in section '4.1 Phases of damaged layer formation'.

**3.2 Structural characterization**

The scanning electron micrograph in Fig. 3A shows a peculiar cellular structure on the surface of the 200-keV $Sb^+$-implanted Ge sample. The surface contains cavities (black appearance) with horizontal dimensions of about 10 nm to 100 nm. Holland et al. regarded this structure as honeycombs [39]. Very similar surface structure has also been observed after 3 MeV $I^+$ irradiation of Ge [43]. SRIM calculation shows a Sb peak atomic concentration of few percent (without taking into account the effect of sputtering and void formation). A damaged range of approximately 100 nm can be anticipated from the SRIM vacancy distribution (Fig. 3B). The peak damage is located at about 40 nm depth and extends to the end of range of the implanted ions. Note that a displacement threshold energy of 30 eV [52] for Ge atoms has been considered in SRIM simulations.

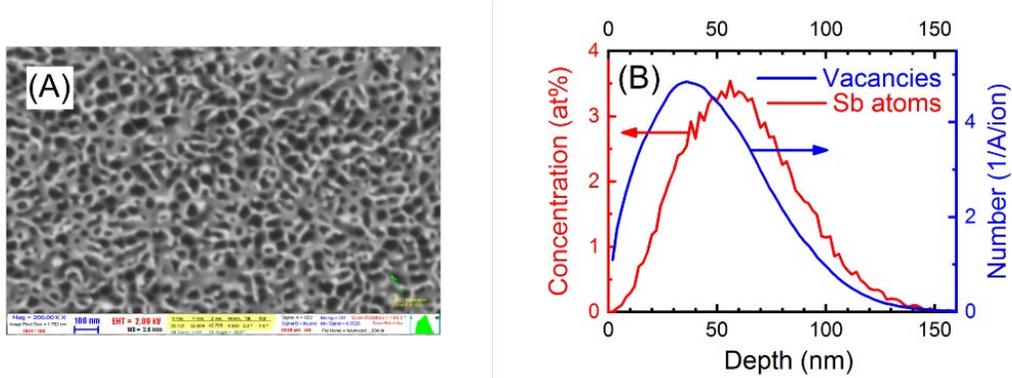

**Figure 3.** (A) SEM micrograph showing the peculiar cellular structure on the surface of c-Ge implanted by 200-keV Sb at a fluence of $1\times10^{16}$ cm$^{-2}$ and ion flux of $2.1\times10^{12}$ cm$^{-2}$s$^{-1}$. (B) Depth distribution of the implanted Sb atoms and vacancies calculated by SRIM [51].

**3.3 Phases of damaged layer formation**

The features of the measured ellipsometric signals (Figs. 2 and 4) as well as the evolution of the implanted structure can be subdivided into 4 regions: (I) a baseline before switching on the ion beam, (II) damage accumulation and amorphization, (III) rapid void formation, (IV) slow change of damage and void constituents, revealed by the increasing value of Ψ in Fig. 4.

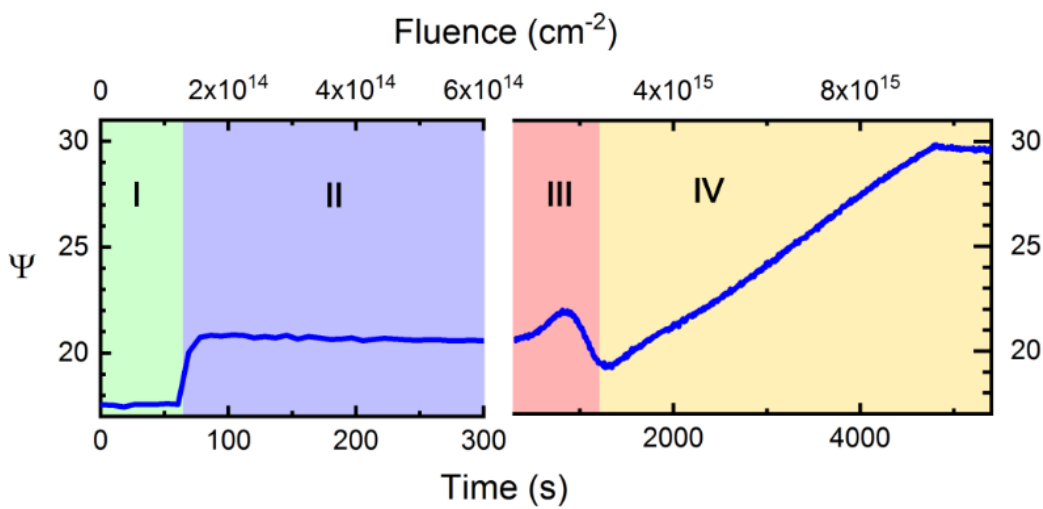

**Figure 4.** Evolution of the ellipsometric angle Ψ measured in-situ at the wavelength of 450 nm during the implantation of 200-keV Sb$^+$ ions into c-Ge. The four dominant regions of structural changes are numbered and marked by different background colors. The regions represent (I) the baseline for the unimplanted case (II) the damage accumulation and amorphization (III) the rapid void formation and (IV) the period of slow change of damage and void constituents. Note the different time for scale of regions (I) and (II) for a better visualization of the transient at ≈60 s, i.e., the starting point of the implantation.

The baseline of the ellipsometric angles was taken from region I and it was used to determine the dielectric function of the c-Ge substrate. From the numerous possible parameterizations of the dielectric function [31], the dispersion of the c-Ge wafer was described using the Johs-Herzinger generalized critical point model [53]. Only the oscillator parameters of Ge transitions at 2.1, 2.3, and 3.4 eV were fitted in a c-Ge/GeO$_2$ model. Subsequently, these parameters were fixed during the evaluation of the spectra measured during the ion implantation.

In regions II-IV a two uniform sublayer model was used. According to the SRIM simulation the modified layer can be divided to two sublayers. The first sublayer can be characterized with increasing dopant and void concentration and the second sublayer with decreasing dopant and void concentration (c.f. Fig. 3B). Both sublayers were built from components of c-Ge, implantation-amorphized Ge (i-a-Ge) and voids. Note that for the i-a-Ge component the parameters were taken from our previous work [20]. The effective dielectric function of each layer was calculated using the Bruggeman EMA [54]. The initial dielectric function of i-a-Ge [20] was parameterized by the Tauc-Lorentz dispersion model, in which only the energy position of the single Lorentz oscillator was fitted. A deviation in the dielectric function of i-a-Ge makes sense, considering that amorphous semiconductors may have different optical properties depending on the preparation conditions [25,26,38,20,32,55,56].

Fig. 5 shows measured and fitted Ψ and Δ spectra for selected characteristic temporal points of structure formation. The curves vary significantly in terms of time and spectral features. Note that the repeatability of the measurements of both Ψ and Δ is ≈0.2°, that is, a small fraction of the width of the plotted lines.

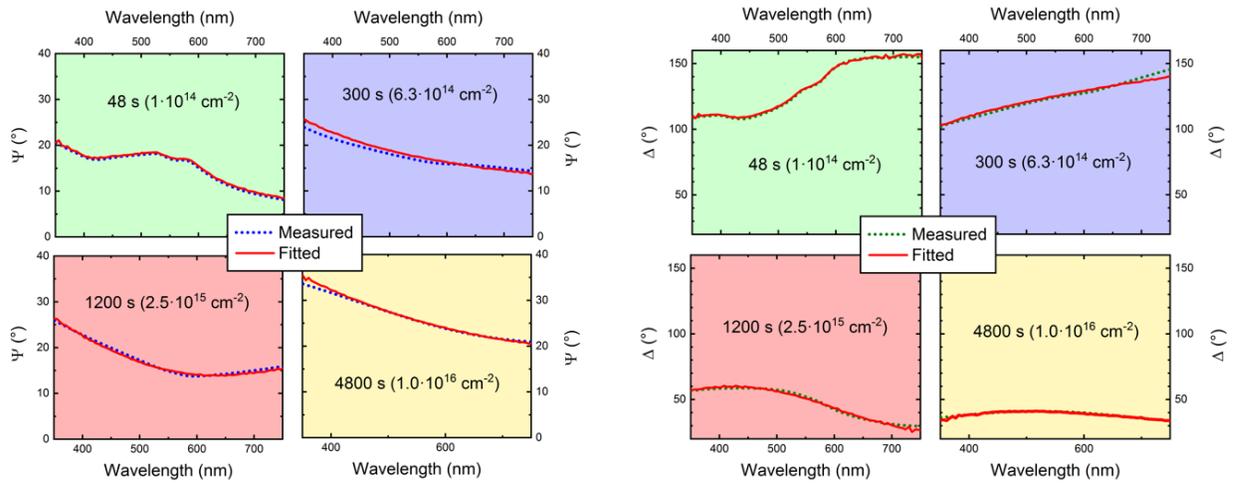

**Figure 5.** Ψ and Δ ellipsometric angles at selected temporal positions corresponding to regions I-IV of Fig. 4 measured in-situ during ion implantation. The measurement errors in both Ψ and Δ are approximately 0.2°. Solid red lines show the spectra calculated from best-fit optical models discussed in the text.

Each SE spectra, collected during Sb implantation, were evaluated according to the four regions I-IV. Let us mention that the full measurement run covers approximately 90 minutes of data acquisition with 3 s time steps, i.e., almost 2000 individually fitted spectra. As a consequence, the fluence of 2x10$^{13}$ is just sufficient to amorphize Ge in the peak damage region (≈40 nm depth). Further implantation leads to gradual amorphization of deeper regions thus appearing as an a-Ge layer with increasing thickness. After the amorphization process higher fluences up to ≈10$^{15}$ cm$^{-2}$, i.e., in the 1-10 dpa range do not cause significant change in SE data. In this fluence range accumulation of point defects and implanted Sb atoms may occur without drastic structural and density changes in the near surface region of Ge, see the smooth OPD values in Fig. 2. In contrast, significant increase of the void fraction and of the OPD can be observed above a fluence of about 10$^{15}$ cm$^{-2}$, see Fig. 6 and Fig, 2. In stage IV significant quasi-periodic structural reorganization may occur in the near surface and/or in deeper regions in 1-2x10$^{15}$ cm$^{-2}$ fluence steps that is reflected in changes of the ellipsometric parameters in Figs. 2, 4, and 6.

The results of SE spectrum evaluation are summarized in Fig. 6. Fig. 6A shows evolution of the thickness of the modified layer, also indicating the thicknesses of the sublayers. The borderline of the sublayers is marked by red color. Furthermore, in Fig. 6A the volume fractions of c-Ge, i-a-Ge and voids in the sublayers are indicated with different shades.

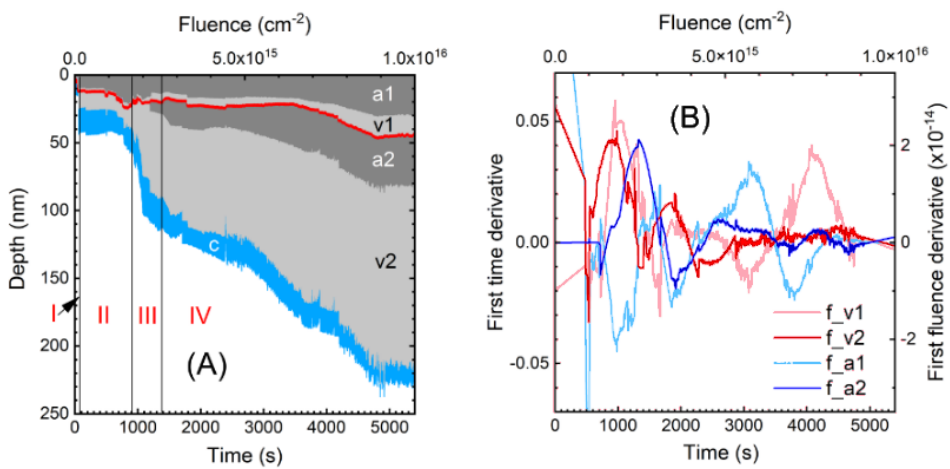

**Figure 6.** (A) Evolution of the SE determined modified layer thickness and sublayer thicknesses (areas separated by a solid red line). The composition of the sublayers is denoted by the volume fraction of the components of i-a-Ge (dark grey areas denoted by 'a'), voids (light grey areas denoted by 'v') and c-Ge (light blue area denoted by 'c') as a function of time. The ratio of the areas covered by the individual components on both sides of the solid red line is proportional to their volume fractions. The fitted parameters were the thicknesses of the sublayers (uncertainty of a few nm) and the volume fractions (uncertainty of a few %) of i-a-Ge and void. The regions of damage formation I-IV are also marked by vertical lines. The white area is the single-crystalline substrate. (B) Time and fluence derivatives d$f_{a1}$/dt, d$f_{a2}$/dt, d$f_{v1}$/dt, and d$f_{v2}$/dt, respectively, of the volumetric fractions of i-a-Ge and voids in the two different sublayers (uncertainty is better than 0.01 for the time derivative).

Although the repeatability of the Ψ and Δ spectra are better than 0.2°, as indicated for Fig. 5, the uncertainty of the fitted parameters is larger, comparable to the noise shown in Fig. 6A at the edge of the stripes representing the approximate uncertainty of the measurement (a few nanometers and a few percent for the thicknesses and the volume fractions, respectively). The thickness of the layers rapidly changes in region III, partly because of the increasing penetration depth of the implanted ions due to the void formation. The void fraction in the top layer accounts for the roughness [54] as well. In the last phase of structure formation (region IV) both the void fraction and the thickness of the bottom layer increase monotonically. Crystalline germanium (c-Ge) can be found only in the second, deeper sublayer with approximately the same content in each region. In Fig. 6B the time / fluence derivatives of the volumetric fractions of voids and i-a-Ge components are shown in the two sublayers.

## 4. Discussion

### 4.1 Time evolution of the surface layer during Sb implantation

The clear advantage of our in-situ SE measurements can be discerned when looking at the time – or fluence - derivatives of the SE parameters, i.e., the volume fractions of void and i-a-Ge components in the different sublayers, see Fig. 6B. These quantities give valuable information on the evolution of the void and damage structure as a function of time (fluence) and help to identify subsequent stages of structural changes and dynamics of spatial reorganization in the material during ion bombardment. Note that usually the effect of ion implantation into Ge is followed by ex-situ characterization of the damaged zone. In such cases the sample is described after performing ion bombardment to certain fluence values [19,41]. Even if the main features of void formation can be followed in this manner, e.g., through the measurement of surface roughness at certain fluences [41], or by TEM analysis which provide information about the extent and growth of voids in three dimensions [41], accelerations and decelerations in the intensity of structural changes vs. irradiation time or vs. the fluence can hardly be deduced from ex-situ studies.

In Fig. 6A, changes in the slope of the effective layer thickness and volumetric fraction curves can be observed as a function of time which suggests certain long-term dynamic character in the void formation process. These changes become more accentuated in Fig. 6B, where the derivatives of volume fractions are shown for voids and i-a-Ge, both for the subsurface and the buried damaged sublayers (layer 1 and layer 2). Clearly, an oscillating character appears for each component with characteristic time (fluence) intervals of 500-1000 s (1-2x10$^{15}$ cm$^{-2}$). To our knowledge, to date no such intermittent or oscillation type behavior was reported for the time dependent void formation process. This finding shows that besides the characteristic fluence required for initiation of intense void formation at about 1x10$^{15}$ cm$^{-2}$ in stage III, other characteristic fluence regions of void formation and amorphization also exist within stage IV. This trend suggests the quasi-periodic atomic scale reorganization of the material.

As Fig. 6B shows, void formation in layer 2 becomes intense at a fluence of ~$2 \times 10^{15}$ cm$^{-2}$ and this is followed by accelerated amorphization peaking at ~$2.5 \times 10^{15}$ cm$^{-2}$. Thereafter, void formation dominates around a fluence of $3.5 \times 10^{15}$ cm$^{-2}$ meanwhile the volume fraction change of i-a-Ge decreases. Another maximum for the i-a-Ge and minimum for the void fraction change can be seen at a fluence of about $5 \times 10^{15}$ cm$^{-2}$. In the thinner near surface region, sublayer 1, the void formation rate is increased at fluences of about $2 \times 10^{15}$ cm$^{-2}$ and $7.7 \times 10^{15}$ cm$^{-2}$ while the amorphization rate is raised in between, at a fluence of $5.7 \times 10^{15}$ cm$^{-2}$, respectively. The rate of volume fraction change of i-a-Ge decreases at fluences of around $2 \times 10^{15}$ cm$^{-2}$, $3.6 \times 10^{15}$ cm$^{-2}$, and $7 \times 10^{15}$ cm$^{-2}$. The first and third minima are about in coincidence with intense void formation stages.

The above-described timeline suggests that the precondition for progressive void formation may be the appearance of newly amorphized regions in the c-Ge substrate which leads to the accumulation of new, freely migrating vacancies and interstitials. After or in parallel with the amorphization step a reorganization process takes place leading to an increase in the total volume of voids probably through the coalescence of smaller vacancy clusters. The amplitude maxima in the time derivatives follow each other by roughly equal steps on the fluence scale within a range of $1-2 \times 10^{15}$ cm$^{-2}$. As a possible scenario, in the first rapid void formation step starting at about $1 \times 10^{15}$ cm$^{-2}$, open volumes may appear in the yet amorphized region thus allowing penetration of newly incoming Sb$^+$ ions into underlying undamaged Ge in order to form newly amorphized zones. These newly amorphized volumes can be transformed to voids through the migration of vacancies and interstitials. We propose this mechanism based on previous works which by TEM analysis reveal the appearance of a considerably ordered columnar void structure oriented perpendicular to the sample surface [41,57]. Also, it was pointed out that the maximum depth of voids can exceed several times the projected range of the implanted ions [19,41,58]. Such deep extension of voids is in agreement with the high diffusion coefficient for vacancies in Ge (~$10^{13}$ cm$^{-2}$ s$^{-1}$) allowing to diffuse as far as ~10 nm within one second at room temperature (RT) [52,59]. For interstitials, however, orders of magnitude lower diffusion lengths can be estimated [52,59].

In the further stages of implantation these two process steps – damage formation and coalescence of vacancy agglomerates - may alternate on the timescale up to the highest fluence applied. The rapidly decreasing amplitudes of oscillations for sublayer 2 in Fig. 6B suggest saturation-like behavior of buried amorphization and void formation processes. However, this trend also can be explained by the fact that the increasing effective thickness of the damaged layer gradually gets closer to the maximum information depth of the SE analysis. In addition, newly introduced amorphous and void zones represent less and less fraction within the effective thickness of sublayer 2. This condition can also be reflected in the decreasing amplitudes of oscillation. To confirm the role of the different factors in the decrease of oscillation amplitudes requires more detailed data analysis with a higher number of sublayers to be introduced in the damaged zone. The subsurface region, i.e., sublayer 1 also shows oscillation like behavior, but in this case an intense decrease of the oscillation amplitudes vs. time cannot be observed in Fig. 6B. Note that previous studies did not show saturation in the depth of damaged zone for 50 keV Ge$^+$ implanted Ge up to a fluence of $2 \times 10^{17}$ cm$^{-2}$ [19], and, also, no saturation of the surface roughness increase vs. ion fluence was reported for high-fluence Sb$^+$ implanted Ge [41].

Note that structural changes in stage IV occur at about two orders of magnitude longer characteristic timescale than the amorphization process in stage I. The difference can be associated with the degree of thermal activation of void formation at RT [60], and/or with the fact that spatial reorganization may occur on significantly longer spatial scale for the movement of defects in stage IV (defect agglomeration) as compared to that in stage I (single atomic displacements). Moreover, distinct features of the void formation process in the different sublayers may be associated with the fact that at a certain time the damage levels and thus the concentration of vacancies and interstitials vary as a function of depth due to the depth dependence of defect generation (dpa) rates during the implantation process.

In summary, Fig. 6 shows that the in-situ SE spectra of this study contain valuable information on the time evolution of void formation in ion implanted Ge. Such results may initiate new experiments to describe and understand reorganization steps in the material through the construction of a detailed model picture of the underlying physical processes.

**4.2 Ion beam-induced amorphization and track size**

The effective size of damaged zones formed from ion tracks initiated by individual bombarding ions can be estimated by numerical simulation compared with the dynamics of damage profile changes measured by RBS and ellipsometry [61] vs. the implanted fluence. In this case the numerical simulation assumes both a track size and a damage profile, the latter can be measured by RBS. The evolution of damage was measured and calculated as a function of ion fluence, i.e., as snapshots of the process in time. In the case of ellipsometry the analysis of the vanishing absorption features [62] was enough to determine the dynamics of amorphization and the track size [61].

A much simpler assumption is to take into account that the ions impinging at random positions of the surface satisfy the Poisson statistics, i.e., the probability that the next ion finds a non-damaged position is proportional to the area of the surface that has not yet been amorphized [62]. It has been shown that an individual ion track generated by a bombarding ion leads to rapid temperature increase and melting of the track zone in the target material [40]. This process results in the formation of a residual circular amorphous zone with its radius on the nanometer scale around the ion trajectory due to fast cooling and quenching of the displaced target atoms in the collision cascade zone. The radius of such residual amorphous tracks depends mainly on the energetics of the collision cascade, i.e., the mass and energy of the implanted ion and the properties of the target material. Therefore, a characteristic ion track radius ($R_T$) can be anticipated for each implantation process with given ion mass, ion energy, and target material [40].

Fig. 7 shows the fit considering Poisson statistics for the evolution of $f_a$, i.e., the amorphous i-a-Ge fraction. The red line corresponds to the Poisson $l$ parameter value of $l \approx 11$ (computed in fluence units of $1 \times 10^{12}$ cm$^{-2}$), and from l a corresponding track diameter of 2.5 nm (track area of 5 nm$^2$) can be estimated. In the calculation it was

14/23

considered that $l = np$, where $n$ is the total number of possible individual events (total number of different independent positions for incoming ions within a window of 1 cm², i.e., the inverse of the track size, $(R_T^2 p)^{-1}$, with $R_T$ being the track radius), and $p$ is the probability to find undamaged region by an incoming ion, for which the center value of $p = f_a = 0.5$ was considered.

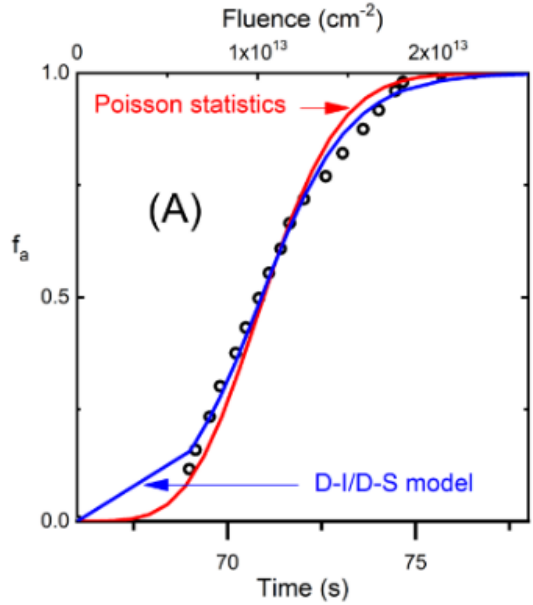

**Figure 7.** Fit of the amorphous fraction i-a-Ge derived from SE measurements (open dots) considering the direct-impact, defect-stimulated (D-I/D-S) amorphization model (blue line) and the Poisson statistics (red line) for implanted Ge. For more details see the text.

To ascertain more details about the progress of implantation-induced damage accumulation, we applied the direct-impact, defect-stimulated (D-I/D-S) amorphization model to reproduce the change in the amorphous fraction i-a-Ge vs. the applied ion fluence. Previously, this model has been successfully applied to describe the behavior of Ge [43] and other semiconductor materials [63] exposed to ion irradiation. In the D-I/D-S model, amorphous nuclei are directly produced in the core of a cascade (homogeneous amorphization) and irradiation-induced point defects and/or subsequently implanted ions stimulate further amorphization at the crystalline-amorphous interfaces (heterogeneous amorphization). If the probability for stimulated amorphization is taken as $f_a(1-f_a)$ then the differential change in $f_a$, due to an infinitesimal fluence, $dD$, can be written as:

$$df_a / dD = s_a(1-f_a) + s_s f_a (1-f_a) \qquad (1)$$

where $s_a$ is the direct-impact amorphization cross-section, and $s_s$ is the effective cross-section for stimulated amorphization [63]. The fit for $f_a$ can also be seen in Fig. 7 with corresponding values of $s_a = 0.6$ nm² and $s_s = 40$ nm², respectively. Note that, while the value of $s_a = 0.6$ nm² is comparable to an effective cross-section for defect formation, $s_{eff} \approx 0.5\text{-}1$ nm² derived from SRIM vacancy profiles, the cross-section $s_s = 40$ nm² is much larger. That is, the overall damage formed in cascade processes initiated by one impinging ion is several times higher than predicted by SRIM. This result is consistent with molecular dynamics (MD) simulations performed for 5 keV $Sb^+$ bombardment into Ge, showing that a large number of defects can be formed in hot collision cascades, and most of them is contained in larger defect clusters, which can be thermodynamically more stable than single point defects [64]. The probability to form such complex defect structure is much higher for heavy $Sb^+$ ions than, e.g., for lighter $B^+$ or $Si^+$ projectiles [52,64]. For 5-keV $Sb^+$ bombardment MD simulations show that the total number of atomic displacements (about 2000/ion) is about 20 times higher than predicted by SRIM (about 100/ion). This result is in agreement with our observations based on the D-I/D-S amorphization model. In general, the high damage cross-section for heterogeneous amorphization in $Sb^+$-implanted Ge suggests its tendency for significant local atomic transport under these conditions that can be a prerequisite to initiate a spatial reorganization process and the formation of voids at higher ion fluences.

The different cross-sections for damage formation, obtained from Poisson statistics and from the D-I/D-S model are due to the distinct basic assumptions applied. Note that the cross-section value given by the geometrical concept-based Poisson function falls between the values of $s_a$ and $s_s$ provided by the D-I/D-S model, which combines contributions to $f_a$ from both homogeneous and heterogeneous amorphization. In the D-I/D-S model the effect of the lower cross-section $s_a$ is compensated with the higher cross-section $s_s$ to reproduce the shape of the amorphous fraction curve, $f_a$.

Our derived cross-sections for direct-impact, and heterogeneous amorphization differ from the values found for 3 MeV $I^+$ ion irradiation, where $s_a = 9$ nm$^2$ and $s_s = 20$ nm$^2$ were reported [43]. The reason for the differences can be explained considering the dissimilar conditions for the lower energy $Sb^+$ and the high energy $I^+$ irradiation. The electronic energy deposition in "hot" collision cascades which are associated with temperature increase and local target melting [40] is about 4 times higher for the $I^+$ irradiation, as predicted by SRIM, and therefore the probability to directly form larger amorphous clusters via local melting and fast cooling [40] is higher compared to our case of $Sb^+$ implantation into Ge. On the other hand, the higher local temperature may be accompanied by different in-situ defect formation, annealing, cluster formation and diffusion kinetics of defects for 3 MeV $I^+$ compared to 200 keV $Sb^+$. It is worth noting that in Ref. [43] the amorphous Ge fraction has been extracted from ex-situ RBS/C measurements while in this work these data were obtained from in-situ SE spectra.

Based on the in-situ method and the models developed in this work, before long we plan to investigate the initial stages of amorphization with lower ion doses, other types of ions and in-situ annealing to identify technologically relevant process parameters that allow to avoid the deteriorating effect of void formation during ion implantation into c-Ge. Note that the dynamics of void and point defect formations can only be analyzed by using in-situ characterizations of this kind with a suitable temporal resolution.

## 4.3 Mechanisms of damage and void formation

The temporal change of the ellipsometric model parameters is shown in Fig. 6A, using 2 layers on intact Ge substrate. Both layers contain c-Ge, i-a-Ge ($f_a$) and void ($f_v$) phases. The volume fraction of void, $f_v$, has previously been estimated in other works by the expansion of the sample [66] measured ex-situ by Talystep or by the analysis of TEM images [41]. Contrarily, in our case the void volume fraction $f_v$ from the ellipsometry fit was measured in real time during irradiation. Due to the presence of voids, the gradient profiles might be more complex than those used for simple unperturbed Gaussian damage profiles [21,24]. The approximate accuracy of the determination of $f_v$ was a few percent, at a time resolution of 3 s. Note, that in the SE model both i-a-Ge and voids are distributed within the modified depth zone continuously. The two homogeneous composite sublayers are a simplification of a depth profile that in the future may better approximated by an analytical gradient profile based on a wider spectral range of SE measurements.

After starting the irradiation (temporal position of ≈60 s), the thickness of the surface layer increases to a value of ≈50 nm within less than 10 s. Note that the penetration depth of light in i-a-Ge in the used wavelength range is a couple of times 10 nm, which means that the saturation of $d_a$ may be caused by the limited penetration of light.

Based on the paper written by Kaiser et al. [41], the critical fluence of void formation for Sb implantation into Ge is between $3x10^{14}$ and $5x10^{14}$ cm$^{-2}$. These values were comparable to that found for Ge self-implantation ($2x10^{15}$ cm$^{-2}$) [42]. In our experiment void formation occurs at a fluence of $1x10^{15}$ cm$^{-2}$ measured with a high accuracy using the in-situ SE method (see Region III in Fig. 4 and also in Figs. 6A and 6B).

The proposed explanations of void formation in the literature range from sputtering and redeposition [19,67] through thermal spikes [68] to clustering of vacancies and diffusion of interstitials [39,60]. Although sputtering might cause removal of surface atoms, it has also been shown by different authors that the void formation obeys mass conservation, and the sputtering effect can be ruled out [19,41,69] in the mechanism of void formation, also proved by the use of a capping layer [57], by molecular dynamic simulations [67] and by scanning tunneling microscopy (STM) measurements [68].

Formation of columnar voids of 20-40 nm in diameter in Ge have been reported by many authors for different experimental (preparation) conditions [19,66,70]. It was found for high-energy heavy ion bombardment that the voids are formed by the agglomeration of vacancies, and a critical defect production rate is necessary to initiate the formation of the sponge-like structure [69]. It was found feasible already in the early investigation of Appleton et al. that the primary cause of void formation is the high mobility of defects [70]. This was underlined by the fact that implantation of 280-keV Bi into Ge at liquid nitrogen temperature (LNT) did not lead to void formation, whereas annealing and high-temperature ion implantation leads to the decrease of dopant retainment, [70] also pointing out the gettering effect of voids [37].

In case of self-implantation at the energy of 500 keV flattened vacancy agglomerates of 10.6 nm average size are formed at a fluence of $1x10^{15}$ cm$^{-2}$ at RT [60]. As shown in Figs. 6 and 7, this is the temporal position in our measurement from which a pronounced growth of the damaged and porous layer starts. The study of Desnica-Fankovic et al. [60] also shows that the size of these agglomerates grows to 17 nm at the fluence of $3\times10^{15}$ cm$^{-2}$, and their shape becomes more spherical. At higher fluences, these nanoclusters agglomerate into larger voids of a broad size distribution, which completely dominates at afluence of $3\times10^{16}$ cm$^{-2}$. The thermal energy in the RT-implanted samples is large enough for the diffusion and restructuring of defects and clustering the vacancies into voids that finally cause porosity – a feature that is lacking in the case of implantation at LNT [60,71].

The voids remain stable during annealing, which restricts shallow junction formation by the implantation of heavy elements [72]. The void formation is the reason for finding unintentional O and C impurities by ion scattering measurement after removing the samples from vacuum [70]. This structure can, however, be used for gettering as well [37,69]. The structure created by the implantation of Sb into Ge can also be reproduced by other elements if the fluence and the temperature is high enough [39,57,73,74]. Therefore, in device fabrication, lighter elements are used as dopants [72]. E.g., B (with Ge pre-amorphization) and P (with self-amorphization) can be used as impurities with subsequent low-temperature annealing for effective dopant activation.

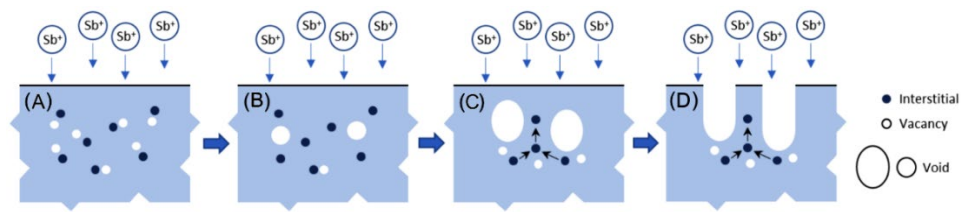

**Figure 8** (A-D) Proposed mechanism of void formation [75,76]. Small dark and bright dots show the interstitials and the vacancies, respectively. The large bright areas are voids. In panel (A) interstitials and vacancies are formed in the initial phase of the implantation, this is followed by the formation of small voids, also in the early stage of the process (B). The interstitials surviving the recombination process migrate together and aggregate in regions between the voids (C). Finally, the voids grow, perpendicular to the surface, and burst so that a cellular structure is formed (D). The appearance of this surface structure is shown in Fig. 3A.

A detailed picture of the void formation mechanisms was shown by Nitta et al. for GaSb implanted by Sn$^+$ [75,76]. In this case, voids are formed by the migration of interstitials in the first stage of implantation (Figs. 8A and 8B, as well as region II in Fig. 5). The interstitials are not stable at RT [77], and those that still survive the annihilation can migrate to the bottom of walls. In this model the walls develop by the aggregated interstitials (region III in Fig. 6A), while the voids by the vacancies migrating to the bottom of existing voids (Fig. 8B). This model does not explain the driving force for the interstitials to aggregate at the bottom of the walls. A plausible background to this phenomenon might be the different depth profiles of interstitials under the voids and the walls between the voids (due to the different amounts of materials above a certain position in depth depending on the amount of void in the path of the penetrating ion), which results in a lateral gradient of interstitial concentration in the lattice.

In the final stage of implantation the voids burst to the surface, as shown in Fig. 8D. These processes are confirmed by the present results which show that the volume fraction of voids is larger in the embedded layers, and the void fraction in the surface layer still increases at the end of the process (see Fig. 6A and the maximum in $df_{v1}/dt$ at about t ≈ 4200 s in Fig. 6B). Note that the damage depth range at the initial stage is consistent with the SRIM ion range calculation shown in Fig. 3B. However, in stage IV, the modified layer with voids and amorphous Ge extends to a depth much greater than the SRIM projected range of Sb+ ions, and this is in agreement with previous observations [19,41].

## 5. Summary and conclusions

An in-situ method to observe the dynamics of structural damage accumulation during ion-implantation was developed and presented. Adverse void formation and evolution of subsurface nanocavities or a cellular surface texture once formed in heavy ion implanted Ge, cannot be removed by annealing. Real-time high temporal resolution in-situ SE measurement combined with an appropriate optical model allow the continuous determination of sample structure-related model parameters and their time evolution during the ion implantation process The concept helps to understand dynamic aspects of damage formation. Quantitative data such as the size of tracks created by the implanted ions, the volume fraction of phases, and the time intervals of quasi-periodic oscillations that can be observed in the three-dimensional reorganization process of the material structure can also be extracted. In-situ SE may be widely applied as a non-destructive technique to understand physical phenomena taking place during ion implantation and to optimize application related technological processes.

**Declaration of competing interest**

The authors declare that they have no known competing financial interests or personal relationships that could have appeared to influence the work reported in this paper.

**Acknowledgements**

The authors are grateful for financial support from the OTKA K131515 and K129009 projects.

**Data availability statement**

Supplementary data associated with this article will be found, in the online version